\documentclass[a4paper,fleqn]{cas-dc}

\usepackage[authoryear,longnamesfirst]{natbib}
\usepackage{amsmath,amssymb}
\usepackage{graphicx}
\usepackage{float}

\ExplSyntaxOn
\bool_gset_true:N \g_stm_nologo_bool

\ExplSyntaxOff

\begin{document}
\let\WriteBookmarks\relax
\def\floatpagepagefraction{1}
\def\textpagefraction{.001}

\shorttitle{Adverse Selection with Quality Variance}
\shortauthors{Z.-L. Zhang, T.-J. Zhou and C.-P. Sun}

\title [mode = title]{Adverse Selection with Quality Variance: A Maximum-Entropy Approach}

\author[1]{Zhi-Lei Zhang}

\author[1]{Tan-Ji Zhou}

\author[1]{C.P. Sun}\cormark[1]
\ead{suncp@gscaep.ac.cn}

\affiliation[1]{organization={Graduate School of China Academy of Engineering Physics},
            city={Beijing},
            citysep={},
            postcode={100193},
            country={China}}

\cortext[1]{Corresponding author.}

\begin{abstract}
The adverse-selection mechanism in markets explains how asymmetric information between buyers and sellers can drive high-quality goods out of the market, thereby causing market deterioration. In its simplest formulation, only the mean quality is used to describe the market, and this is insufficient to determine how fast the market deteriorates or how the quality distribution evolves. To resolve the two problems, we describe the adverse selection as a dynamic truncation of the quality distribution: buyers set an upper bound proportional to the mean quality by a rate $\xi$ that is larger than unity, and sellers whose quality exceeds this upper bound reject an offer and exit the market. The retained market is then characterized by the conditional distribution obtained after this truncation, and the corresponding evolution process is iterated until market quality reaches a stable state. This statistical approach gives three results. (i) We identify a mechanism for preventing complete adverse selection, defined as the process where the quality of the market is driven down to the minimum quality floor. (ii) A larger quality variance or a smaller price premium, defined as the amount by which the payment upper bound exceeds the current mean quality, raises the upper bound on the deterioration in mean quality. (iii) A maximum-entropy benchmark shows numerically how quality variance and the payment rate jointly determine market deterioration and the final stable quality platform. This approach also clarifies how market interventions can slow adverse selection: they may raise buyers’ payment rate, reduce quality variance, or increase the minimum quality floor.
\end{abstract}


\maketitle


\section{Introduction}\label{sec:introduction}

In markets, the adverse-selection mechanism explains how asymmetric information between buyers and sellers can drive high-quality goods out of the market, thereby causing market deterioration. The asymmetry appears in the information on the quality of the product when sellers know the quality of their own goods, whereas buyers observe only the mean quality in the market. Markets characterized by information asymmetry are also referred to as `lemons markets', following Akerlof's used-car example \citep{Akerlof1970}, since a lemon or a used car may appear satisfactory while concealing serious defects. In this case, buyers cannot observe individual product quality and pay only according to the mean quality of goods offered for sale; high-quality sellers may exit, the mean quality falls, and the market may unravel. 
%In this paper, mean quality denotes the first moment of the quality distribution among goods that remain in the market.

In its simplest formulation, however, the market state is summarized only by the mean quality. This description is incomplete because two markets can have the same mean quality but different variances, upper tails, or concentrations of quality near a minimum quality floor. They may therefore follow different adverse-selection paths even when buyers assign the same price to the same mean quality. As a result, the simplest case cannot determine how fast the market deteriorates, how the quality distribution evolves, or which quality platform the market reaches when adverse selection stops.

The lemons market literature has developed the adverse-selection mechanism in several directions. Signaling and screening models study how informed agents transmit private information or how uninformed agents induce different types to reveal themselves through costly actions or contracts \citep{Spence1973,RothschildStiglitz1976}; see \citet{Riley2001} for a review. Other studies examine adverse selection under competitive search, public news arrival, decentralized trade, competition, price transparency, and information design \citep{GuerrieriShimerWright2010,DaleyGreen2012,CamargoLester2014,BochetSiegenthaler2021,KartikZhong2023,DiGiovanPaoloHigueras2025}. A related literature studies quality disclosure, certification, warranties, reputation, minimum quality standards, and certifiable information as responses to hidden quality \citep{Grossman1981,Milgrom1981,DranoveJin2010,DaughetyReinganum2008,BarIsaacTadelis2008,BuehlerSchuett2014,Schmitz2021,GanLi2024,Chopra2024}. These studies analyze how information transmission, equilibrium incentives, and market institutions affect prices, participation, and quality choices. However, these studies do not directly characterize how the quality distribution evolves during the adverse-selection process.

To overcome the limitation of describing the market solely by mean quality, we characterize the adverse selection in detail as a dynamic truncation of the quality distribution. Buyers set an upper bound proportional to the mean quality with a fixed payment rate $\xi>1$. Sellers whose quality exceeds this upper bound reject an offer and exit the market. The retained market is characterized by the conditional distribution obtained after this truncation, and the process is iterated until the mean quality reaches a stable state. This construction turns the adverse selection into a simple moving upper-bound recursion.

The analysis of this approach gives three results. First, in a distribution-independent way, the approach identifies the critical payment rate required to prevent complete adverse selection, defined as the process in which market quality is driven down to the minimum quality floor. When market regulation imposes a positive minimum quality floor, a stable platform of quality above this floor exists only if the payment rate exceeds the critical value $\xi_c = 1$. Such a positive minimum quality floor may arise from market regulation \citep{BuehlerSchuett2014}. Without this positive floor, the corresponding critical value is $\xi_c = 2$, and this is consistent with the market-collapse mechanism of \citet{Akerlof1970}. This shows that our approach identifies a mechanism for preventing complete adverse selection. Second, quality variance and the price premium—the amount by which the payment upper bound exceeds the current mean quality—jointly determine the speed of market deterioration. For a fixed price premium, a larger quality variance raises the upper bound on the deterioration in mean quality; for a fixed quality variance, a larger price premium slows it. Third, the maximum-entropy principle provides specific quality distributions as the benchmark. With only the first moment imposed, the distribution is described by the Boltzmann distribution; with both the first and second moments imposed, it becomes a truncated Gaussian. These benchmark distributions are then used to illustrate how quality variance and the payment upper bound jointly determine market deterioration and the final stable platform.

% The paper is organized as follows. Section \ref{sec:model} defines the truncation recursion and derives a distribution-independent bound for the speed of deterioration. Section \ref{sec:upper bound} studies the minimum payment upper bound and the stable platform near the minimum quality floor. Section \ref{sec:maxent} introduces the maximum-entropy benchmark and shows how variance changes the platform. Section \ref{sec:conclusion} concludes.

\section{A dynamic truncation approach}\label{sec:model}

We study adverse selection as a sequence of markets indexed by trading round \(t=0,1,2,\ldots\), where each round corresponds to one step of quality truncation. The quality of goods is denoted by \(q\) and is measured in monetary-equivalent units. The minimum quality floor is \(q_0\geq0\). Let \(f(q)\) be the initial density of product quality on \([q_0,\infty)\).

Due to asymmetric information, sellers know the quality of their own products exactly, whereas buyers cannot observe the quality of individual products and base their decisions only on the mean quality of the products offered for sale. Therefore, at round \(t\), they set a payment upper bound $b_t$
\begin{equation}
    b_t=\xi \bar{q}_t,\qquad \xi>0 ,
    \label{eq:upper bound}
\end{equation}
where \(\xi\) is the payment rate and $\bar{q}_t$ is the mean quality of round $t$ according to the distribution. The good of quality \(q\) remains in the market only if \(q\leq b_t\). Thus, a larger \(\xi\) means that buyers are willing to support a higher payment upper bound for the same mean quality.

Along a monotone deterioration path $\bar{q}_{t+1}<\bar{q}_t$, the payment upper bounds move downward. The market after round \(t\) can be equivalently described by truncating the initial quality distribution \(f(q)\) to the interval \([q_0,b_t]\). Hence, the mean quality in round \(t+1\) is the conditional mean over this retained interval:

\begin{equation}
    \bar{q}_{t+1}
    =
    \frac{
        \int_{q_0}^{b_t} q f(q)\,dq
    }{
        \int_{q_0}^{b_t} f(q)\,dq
    } .
    \label{eq:direct-recursion-b}
\end{equation}
This is the law of the evolution of the adverse-selection process. A stable market quality is a fixed point of this recursion. Complete adverse selection means the case where market quality is driven down to the minimum quality floor \(q_0\).

The retained-market variance at round \(t+1\) is
\begin{equation}
\sigma_{t+1}^2
=
\frac{
\int_{q_0}^{b_{t}}(q-\bar q_{t+1})^2f(q)\,dq
}{
\int_{q_0}^{b_{t}}f(q)\,dq
}.
\label{eq:variance-direct}
\end{equation}
This variance measures how strongly individual qualities fluctuate around the current mean quality within the retained market.

Eq. ~\eqref{eq:direct-recursion-b} computes the mean quality in the next round by directly truncating the initial quality distribution at the current payment upper bound. This procedure is equivalent to updating the retained distribution round by round and truncating it at the current payment upper bound in each round
\begin{equation}
    f_{t+1}(q)
    =
    \frac{
        f_t(q)\mathbf{1}_{\{q_0\leq q\leq b_t\}}
    }{
        \int_{q_0}^{b_t} f_t(x)\,dx
    } .
    \label{eq:update-density-app}
\end{equation}
Consequently, truncating the current retained distribution at the new upper bound and then computing the next mean yields the same expression, because the retained intervals are nested along a monotone deterioration path. Appendix~\ref{app:sequential} provides the proof. Therefore, in the following analysis, we use only the direct-truncation expression in Eqs.~\eqref{eq:direct-recursion-b} and \eqref{eq:variance-direct}. Together, these equations characterize the evolution of mean quality and the quality variance of the retained market under successive payment upper bounds. We first identify how quality variance and the payment upper bound determine the speed of market deterioration. Then we study the local condition under which the market avoids complete adverse selection and converges to a stable quality platform above the minimum quality floor. This identifies a mechanism for preventing complete adverse selection.

\medskip
\noindent\textbf{Lemma 1 (upper-tail loss identity).}
Suppose \(f\) has a finite first moment and \(\bar{q}_{t+1}<\bar{q}_{t}\). Then
\begin{equation}
\bar{q}_t-\bar{q}_{t+1}
=\frac{\int_{b_t}^{b_{t-1}}(q-\bar{q}_t)f(q)\,dq}
{\int_{q_0}^{b_t}f(q)\,dq} .
\label{eq:tail-identity}
\end{equation}

\medskip
\noindent\textit{Proof.}
By the definition of \(\bar q_t\), it is shown that
\begin{equation}
\int_{q_0}^{b_{t-1}}(q-\bar{q}_t)f(q)\,dq=0 .
\label{eq:centered-zero}
\end{equation}
The integration interval can be split as
\[
[q_0,b_{t-1}]
=
[q_0,b_t]\cup(b_t,b_{t-1}]
\]
which gives
\begin{equation}
\int_{b_t}^{b_{t-1}}(q-\bar{q}_t)f(q)\,dq
=
\int_{q_0}^{b_t}(\bar{q}_t-q)f(q)\,dq .
\label{eq:split}
\end{equation}
Through \eqref{eq:direct-recursion-b} and above equation, this lemma can be proved.\(\square\)

Lemma~1 shows that the one-period decline in mean quality is determined by the total excess quality of the sellers excluded when the payment upper bound falls from \(b_{t-1}\) to \(b_t\), relative to the mass of goods that remains in the market. Hence, a larger loss from the excluded upper tail or a smaller retained market leads to a larger decline in the next-round mean quality.

We next give a bound that connects the speed of deterioration to quality variance and the payment upper bound. Define
\begin{equation}
a_t\equiv b_t-\bar{q}_t=(\xi-1)\bar{q}_t .
\label{eq:price premium}
\end{equation}
For \(\xi>1\), $a_t$ is the price premium which represents the distance between the payment upper bound and the mean quality.

\medskip
\noindent\textbf{Proposition 1 (variance, upper bound, and deterioration speed).}
Suppose \(\bar{q}_t>0\), \(\xi>1\), \(\bar{q}_{t+1}<\bar{q}_{t}\), and the retained-market variance \(\sigma_t^2\) defined by Eq.~\eqref{eq:variance-direct} is finite. Then
\begin{equation}
0\leq \bar{q}_t-\bar{q}_{t+1}
<
\frac{\sigma_t^2}{a_t}
=
\frac{\sigma_t^2}{(\xi-1)\bar{q}_t}.
\label{eq:tail-bound}
\end{equation}

\medskip
\noindent\textit{Proof.}
Applying the Cauchy--Schwarz inequality separately to the retained interval \([q_0,b_t]\) and the excluded upper tail \([b_t,b_{t-1}]\), and using Eqs.~\eqref{eq:variance-direct} and ~\eqref{eq:split}, it gives
\begin{equation}
\begin{aligned}
\sigma_t^2
&=
\frac{
\int_{q_0}^{b_t}
(q-\bar{q}_t)^2f(q)\,dq
+
\int_{b_t}^{b_{t-1}}
(q-\bar{q}_t)^2f(q)\,dq
}{
\int_{q_0}^{b_{t-1}}f(q)\,dq
}
\\
&\geq
\frac{
\left[
\int_{b_t}^{b_{t-1}}
(q-\bar{q}_t)f(q)\,dq
\right]^2
}{
\left[
\int_{q_0}^{b_t}f(q)\,dq
\right]
\left[
\int_{b_t}^{b_{t-1}}f(q)\,dq
\right]
}.
\end{aligned}
\label{eq:variance-decomposition-bound}
\end{equation}
Since \(q-\bar{q}_t> a_t\) on the excluded upper tail,
\begin{equation}
\int_{b_t}^{b_{t-1}}
(q-\bar{q}_t)f(q)\,dq
>
a_t
\int_{b_t}^{b_{t-1}}f(q)\,dq.
\label{eq:upper-tail-price premium}
\end{equation}
Combining Eqs.~\eqref{eq:variance-decomposition-bound} and
\eqref{eq:upper-tail-price premium} yields
\begin{equation}
\sigma_t^2
>
a_t
\frac{
\int_{b_t}^{b_{t-1}}
(q-\bar{q}_t)f(q)\,dq
}{
\int_{q_0}^{b_t}f(q)\,dq
}.
\end{equation}
By Lemma~1, the fraction on the right-hand side equals
\(\bar{q}_t-\bar{q}_{t+1}\). Therefore,
\begin{equation}
\sigma_t^2
>
a_t(\bar{q}_t-\bar{q}_{t+1}),
\end{equation}
which gives the upper bound in Eq.~\eqref{eq:tail-bound}. The lower bound follows from Lemma~1 because the integrand on the excluded upper tail is nonnegative. \(\square\)

Equation~\eqref{eq:tail-bound} shows that the upper bound on the one-period decline in mean quality increases with quality variance and decreases with the price premium. Thus, a larger variance or a smaller price premium permits a larger one-period deterioration.

\section{Critical payment rate for preventing complete adverse selection}\label{sec:upper bound}

We now study the end of the adverse-selection process near a positive minimum quality floor. If \(\bar{q}_t=q_0+\epsilon_t\), where \(\epsilon_t\geq0\) is the excess mean quality above the floor, the retained interval above the floor is
\begin{equation}
h_t=b_t-q_0=(\xi-1)q_0+\xi\epsilon_t .
\label{eq:h}
\end{equation}
Assume that \(f\) is smooth near \(q_0\) and
\begin{equation}
f(q_0+x)=f(q_0)+f'(q_0)x+O(x^2),\qquad x\geq0 .
\label{eq:local-density}
\end{equation}
Then the evolution of $\epsilon_{t+1}$ is
\begin{equation}
\epsilon_{t+1}=\frac{h_t}{2}+
\kappa h_t^2+O(h_t^3),
\label{eq:eps-expansion}
\end{equation}
where $\kappa={f'(q_0)}/{(12f(q_0))}$ represents local density slope. Keeping terms up to second order gives the local evolution
\begin{equation}
\epsilon_{t+1}=\frac{\xi-1}{2}q_0+\frac{\xi}{2}\epsilon_t
+\kappa[(\xi-1)q_0+\xi\epsilon_t]^2 .
\label{eq:local-map}
\end{equation}
The above local expansion requires \(0<h_t\ll1\).

\medskip
\noindent\textbf{Proposition 2 (critical payment rate).}
Suppose \(q_0>0\) and the local approximation \eqref{eq:local-map} is valid. If \(\xi\leq1\), complete adverse selection is locally absorbing at the floor. If \(1<\xi<2\), there exists a locally stable quality platform above the floor for all \(\kappa\) sufficiently close to zero. When \(\kappa=0\), the quality distribution is locally flat near the minimum quality floor. Let \(\epsilon_P\) denote the fixed point of the excess mean quality \(\epsilon_t\), and the corresponding stable quality platform is \(q_P=q_0+\epsilon_P\). Then
\begin{equation}
q_P=q_0+\epsilon_P=\frac{q_0}{2-\xi},
\qquad
\epsilon_P=\frac{(\xi-1)q_0}{2-\xi}.
\label{eq:platform-zero}
\end{equation}
Thus, the critical payment rate needed to prevent complete adverse selection near a positive floor is \(\xi_c=1\). If \(q_0=0\), the critical payment rate becomes \(\xi_c=2\).

\medskip
\noindent\textit{Proof.}
The fix point of Eq. \eqref{eq:local-map} satisfies $\epsilon_t = \epsilon_{t+1} = \epsilon_P$, therefore
\begin{equation}
    \epsilon_P=\frac{(\xi-1)q_0}{2-\xi}. \label{stable_floor}
\end{equation}
When \(\xi\leq1\), the excess mean quality shows \(\epsilon_P\leq0\). However, $q_0$ denotes the minimum quality of goods in the market. Hence, the market quality converges to the minimum quality floor, so the market collapses completely and complete adverse selection occurs. When \(1<\xi<2\), the local evolution is
\begin{equation}
\epsilon_{t+1}=\frac{\xi-1}{2}q_0+\frac{\xi}{2}\epsilon_t .
\label{eq:linear-local}
\end{equation}
Its fixed point is \eqref{stable_floor}, which is positive because \(\xi>1\). Since \(\xi/2<1\) in Eq. \eqref{eq:linear-local}, the fixed point is locally stable. For sufficiently small \(|\kappa|\), the implicit-function theorem and continuity of the derivative preserve the nearby stable branch. If \(q_0=0\), the leading term is \(\epsilon_{t+1}=(\xi/2)\epsilon_t+O(\epsilon_t^2)\), the floor is locally stable for \(\xi<2\). In this case, market quality can converge to the lowest quality level, and complete adverse selection may occur. Complete adverse selection is avoided only when \(\xi>2\). \(\square\)

Proposition 2 gives the first main message. When market regulation or similar mechanisms impose a positive minimum quality floor, complete adverse selection can be avoided once the payment upper bound is slightly above the mean quality, i.e., \(\xi>1\). In contrast, without a positive minimum quality floor, the payment upper bound must exceed twice the mean quality, namely \(\xi>2\), to prevent complete adverse selection. This comparison shows that avoiding market collapse requires two complementary channels. Buyers need to support a sufficiently high payment upper bound relative to mean quality, while regulators can reduce the required payment burden by imposing a positive minimum quality floor.

We now examine how the above conclusion changes when \(\kappa\neq0\). For this purpose, we record how the stable platform depends on the local density slope. Let \(\epsilon_P(\kappa)\) denote the stable fixed-point branch of the excess mean quality. Differentiating the fixed-point condition for \eqref{eq:local-map} gives
\begin{equation}
\frac{d\epsilon_P}{d\kappa}
=
\frac{[(\xi-1)q_0+\xi\epsilon_P]^2}{1-F'(\epsilon_P)},
\label{eq:deps-dkappa}
\end{equation}
where \(F(\epsilon)\) denotes the right-hand side of the local evolution \eqref{eq:local-map}. The denominator is positive on the stable branch. To see this, local stability of the fixed point requires \(|F'(\epsilon_P)|<1\). Therefore, the denominator is positive. Since the numerator in Eq.~\eqref{eq:deps-dkappa} is positive, we obtain \(d\epsilon_P/d\kappa>0\). Thus, a larger local density slope raises the stable platform, while a smaller local density slope lowers it. The above results are distribution-free because they do not depend on the specific functional form of the quality distribution. In the next section, we use the maximum-entropy principle to determine benchmark quality distributions. These distributions are then used to illustrate the economic implications derived above: first, the critical value of the payment rate \(\xi\) depends on whether a positive minimum quality floor exists; second, the local density slope is inversely related to quality variance, so a larger variance lowers the final stable platform.

\begin{figure}
  \centering
  \includegraphics[width=0.88\columnwidth]{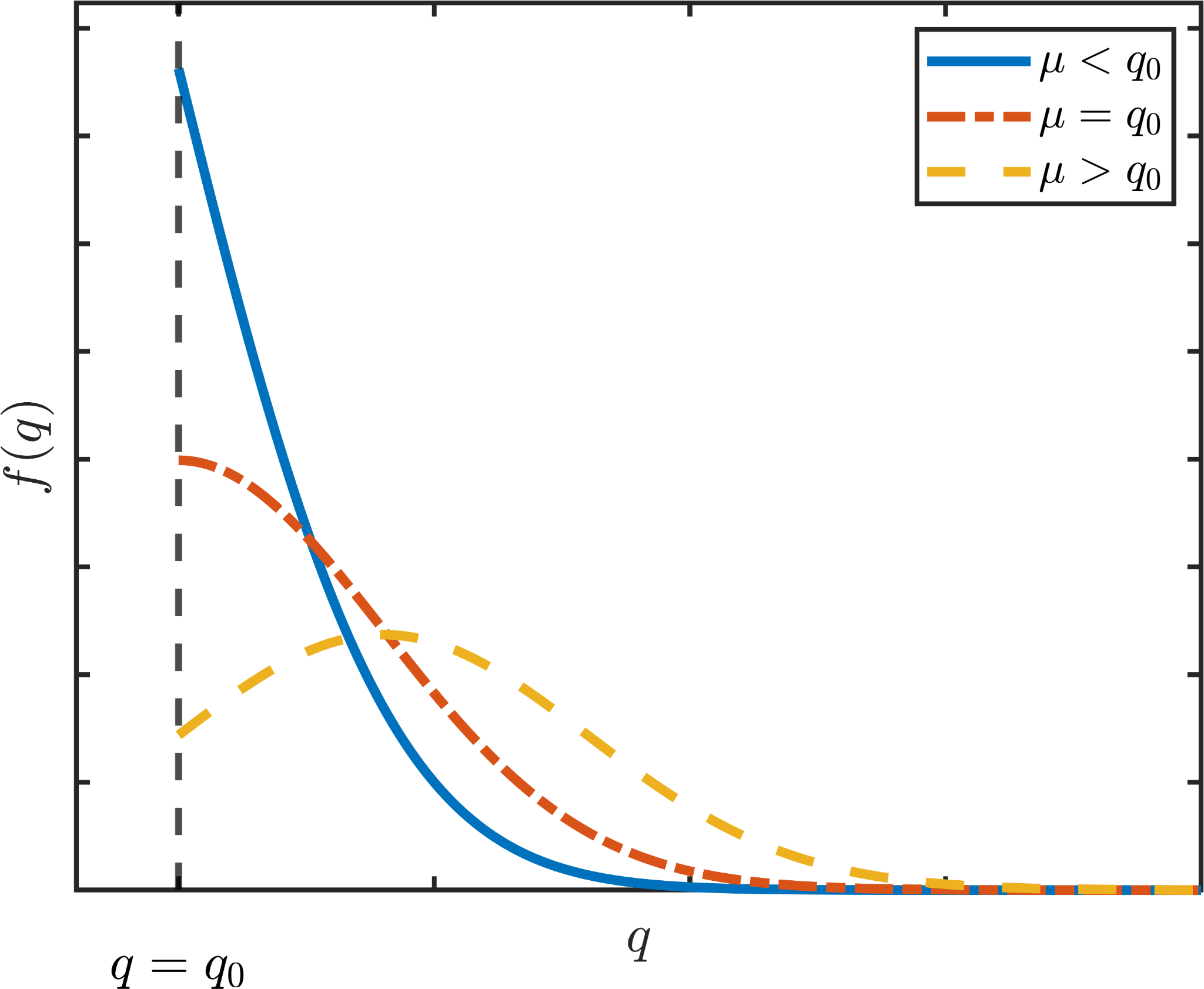}
  \caption{Truncated-Gaussian benchmark distributions for different positions of \(\mu\) relative to the minimum quality floor \(q_0\). The three cases show how the local mass near the floor changes when the same support is used.}
  \label{fig:truncated-gaussian}
\end{figure}

\section{Maximum-entropy benchmark and quality distribution}\label{sec:maxent}

In this section, we use the maximum-entropy principle to determine distributions of quality as the benchmark of our approach. Maximum entropy selects the density with the largest Shannon entropy among all densities satisfying the imposed constraints \citep{Jaynes1957a,Jaynes1957b}. It therefore provides the least informative benchmark consistent with the chosen constraints; see \citet{ScharfenakerYang2020} for economic applications. These benchmark distributions are used to show the economic implications derived above: first, the critical value of the payment rate \(\xi\) depends on whether a positive minimum quality floor exists; second, the local density slope is inversely related to quality variance, so a larger variance lowers the final stable platform.

Let the quality density \(f(q)\) be supported on \([q_0,\infty)\). The Shannon entropy is
\begin{equation}
S[f]=-\int_{q_0}^{\infty}f(q)\ln f(q)\,dq .
\label{eq:entropy}
\end{equation}
If normalization and mean quality \(\bar q\) are the only constraints, the maximum-entropy principle shows that the quality distribution is the Boltzmann distribution
\begin{equation}
f(q)=\beta e^{-\beta(q-q_0)},\qquad q\geq q_0,
\label{eq:exponential}
\end{equation}
where \(\bar q=q_0+1/\beta\) and \({\rm Var}(q)=1/\beta^2\). A mean constraint alone therefore, fixes variance once the mean is fixed. To vary the mean and variance separately, impose the second moment
\begin{equation}
\int_{q_0}^{\infty}q^2 f(q)\,dq=\bar q^2+\sigma^2 .
\end{equation}
The first-order condition gives
\begin{equation}
f(q)=\frac{\exp(-\beta q-\gamma q^2)}
{\int_{q_0}^{\infty}\exp(-\beta x-\gamma x^2)\,dx},
\qquad q\geq q_0 .
\label{eq:quadratic-exponential}
\end{equation}
The parameters \(\beta\) and \(\gamma\) are determined by the imposed mean and second moment. For \(\gamma>0\), define the mean value and variance parameter of the underlying Gaussian kernel by
\begin{equation}
s^2\equiv\frac{1}{2\gamma},
\qquad
\mu\equiv-\frac{\beta}{2\gamma},
\end{equation}
it is shown that the lower-truncated Gaussian density is
\begin{equation}
f(q)=
\frac{1}{s\left[1-\Phi\left(\frac{q_0-\mu}{s}\right)\right]}
\phi\!\left(\frac{q-\mu}{s}\right),
\qquad q\geq q_0,
\label{eq:truncated-gaussian}
\end{equation}
where \(\phi(x)\) and \(\Phi(x)\) are the standard normal density and distribution function, respectively. Here, \(\mu\) and \(s^2\) are the mean value and variance parameters of the underlying untruncated Gaussian distribution. They generally differ from, but are related to, the mean \(\bar{q}\) and variance \(\sigma^2\) of the truncated quality distribution. Fig.~\ref{fig:truncated-gaussian} shows the truncated-Gaussian benchmark distributions for different positions of \(\mu\) relative to the minimum quality floor \(q_0\). The blue solid, orange dash-dotted, and yellow dashed curves correspond to \(\mu<q_0\), \(\mu=q_0\), and \(\mu>q_0\), respectively.

In this case,
\begin{equation}
\frac{f'(q)}{f(q)}=-\frac{q-\mu}{s^2},
\qquad
\kappa=\frac{1}{12}\frac{\mu-q_0}{s^2} .
\label{eq:kappa-tg}
\end{equation}
Thus, holding \(\mu\) and \(q_0\) fixed with \(\mu>q_0\), a larger variance parameter \(s^2\) lowers \(\kappa\). Since \(d\epsilon_P/d\kappa>0\), it lowers the stable platform of the quality.

\medskip
\noindent\textbf{Proposition 3 (variance parameter lowers the platform in the benchmark).}
Under the truncated-Gaussian benchmark, hold the payment rate \(\xi\), the minimum quality floor \(q_0\), and the location parameter \(\mu>q_0\) fixed. Along the locally stable branch, increasing the underlying Gaussian variance parameter \(s^2\) lowers the stable quality platform.

\medskip
\noindent\textit{Proof.}
Equation \eqref{eq:kappa-tg} gives
\begin{equation}
\frac{\partial\kappa}{\partial s^2}
=
-\frac{\mu-q_0}{12s^4}<0 .
\end{equation}
Equation \eqref{eq:deps-dkappa} gives \(d\epsilon_P/d\kappa>0\) on the stable branch. Hence \(d\epsilon_P/ds^2<0\). Since \(q_P=q_0+\epsilon_P\), the platform falls when \(s^2\) increases. \(\square\)

\begin{figure}
  \centering
  \includegraphics[width=0.95\columnwidth]{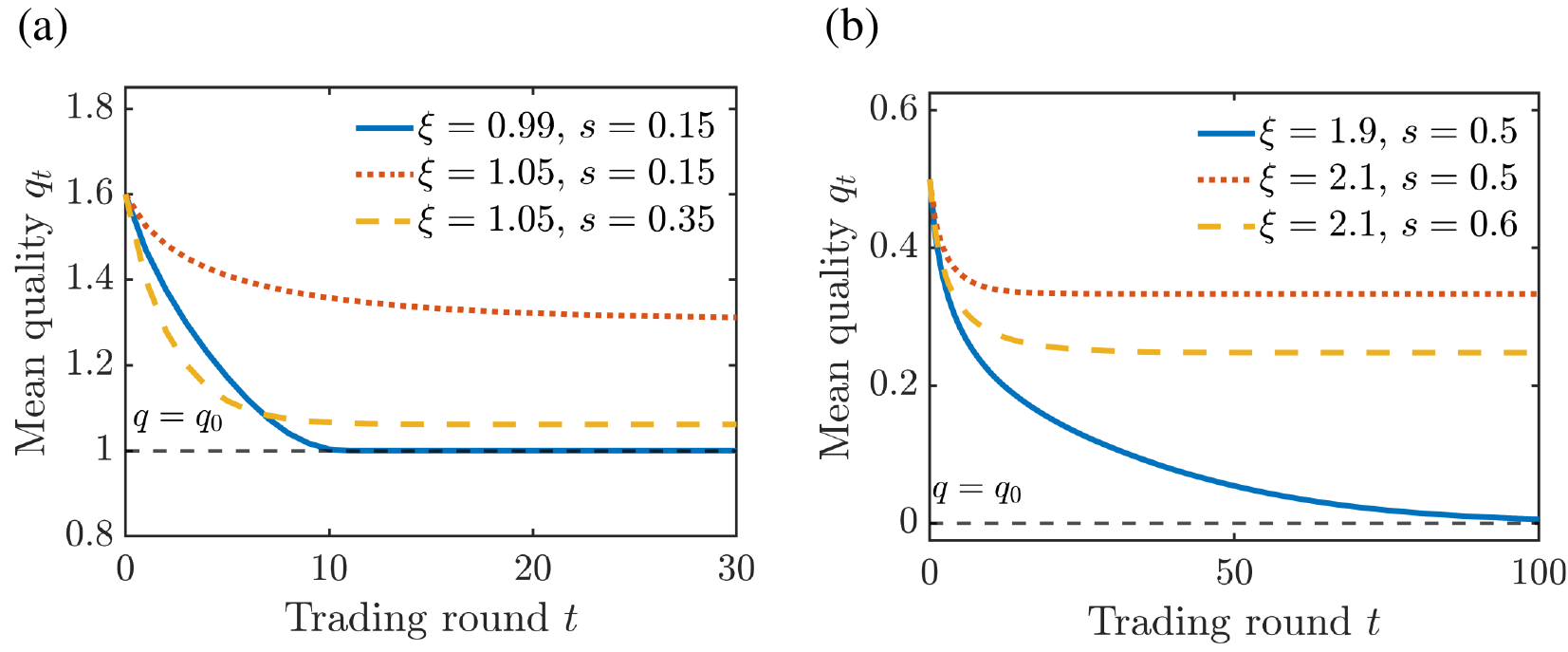}
  \caption{Mean-quality dynamics under the truncated-Gaussian benchmark. 
Panel (a) shows the case with a positive minimum quality floor, \(q_0=1\), initial mean quality \(\bar q_0=1.60\). 
Panel (b) shows the case without a positive floor, \(q_0=0\), initial mean quality \(\bar q_0=0.50\).}
  \label{fig:mean-dynamics}
\end{figure}

Fig.~\ref{fig:mean-dynamics} gives a numerical illustration of the upper bound condition for avoiding complete adverse selection and the effect of the variance parameter on the final stable platform. Panel (a) shows the case with a positive minimum quality floor. 
When the payment rate is below the critical value \(\xi_c=1\) (blue solid line), the mean quality is driven toward the floor; when the payment rate exceeds the critical value, the market converges to a larger stable platform than $q_0$ (see red dotted line and yellow dashed line). Comparing the two curves with the same payment rate but different Gaussian variance parameters ($s=0.15$ for the dotted line and $s=0.35$ for the dashed line) shows that a larger variance parameter lowers the final platform. This is consistent with the local relation \(\kappa=(\mu-q_0)/(12s^2)\): holding \(\mu>q_0\) fixed, a larger \(s^2\) reduces the local density slope, and since \(d\epsilon_P/d\kappa>0\), the stable platform becomes lower. Panel (b) shows the case without a positive floor. The curve with \(\xi<2\) converges toward the lowest quality level, so complete adverse selection occurs, whereas the curve with \(\xi>2\) avoids this local collapse. Economically, the figure shows two points: a positive minimum quality floor lowers the critical payment rate needed to prevent complete adverse selection, and a larger variance parameter weakens the final market quality by lowering the stable mean-quality platform.

\section{Conclusion}\label{sec:conclusion}
This paper has represented adverse selection as a statistical truncation of the quality distribution. Buyers set a payment upper bound according to the current mean quality, sellers whose quality exceeds that upper bound exit, and the retained market is described by the conditional distribution after truncation. Iterating this process determines the path of market deterioration and the stable mean-quality platform reached after adverse selection stops.

The analysis of this approach gives three conclusions. First, the price required to avoid complete adverse selection depends on the existence of a positive minimum quality floor. With a positive floor, complete adverse selection can be prevented as long as buyers are willing to pay a price slightly above the mean quality, i.e., \(\xi>1\). Without such a floor, preventing complete adverse selection requires buyers to pay more than twice the mean quality, that is, \(\xi>2\), consistent with the market-collapse mechanism of the lemons market \citep{Akerlof1970}. Second, the variance of the quality distribution and the price premium, defined as the amount by which the payment upper bound exceeds the mean quality, affect the speed of market deterioration. A larger variance or a smaller price premium allows a larger one-round loss of mean quality. Third, the maximum-entropy benchmark shows how the payment rate and quality variance jointly shape the deterioration path and the final platform.

The economic implication is direct. A market is more vulnerable to deterioration when quality variance is large, and buyers' payment upper bound is low. Market interventions can slow adverse selection by raising the payment upper bound, reducing quality variance, or increasing the minimum quality floor. This provides a distributional description of how a lemons market can either collapse to the lowest quality level or stabilize at a positive mean-quality platform.

%\section*{Data availability statement}
%This is a theoretical paper. No external data or computational code are used.

\appendix

\section{Sequential updating and direct truncation}\label{app:sequential}

Let \(b_t=\xi \bar{q}_t\) be the payment upper bound at round \(t\). Suppose the deterioration path is monotone, so that
\begin{equation}
    b_{t+1}<b_t .
\end{equation}
Then the retained intervals are nested:
\begin{equation}
    [q_0,b_{t+1}]\subset [q_0,b_t]\subset [q_0,b_{t-1}]\subset\cdots .
\end{equation}

Starting from the initial density \(f_0=f\), the sequential updating rule is
\begin{equation}
    f_{t+1}(q)
    =
    \frac{
        f_t(q)\mathbf{1}_{\{q_0\leq q\leq b_t\}}
    }{
        \int_{q_0}^{b_t} f_t(x)\,dx
    } .
    \label{eq:update-density-app}
\end{equation}
We show that this sequential update is equivalent to direct truncation of the initial density. Since the upper bounds are nested, repeated truncation keeps only the smallest retained interval. Hence, by induction,
\begin{equation}
    f_t(q)
    =
    \frac{
        f_0(q)\mathbf{1}_{\{q_0\leq q\leq b_{t-1}\}}
    }{
        \int_{q_0}^{b_{t-1}}f_0(x)\,dx
    } .
    \label{eq:direct-density-app}
\end{equation}
It is shown that Eq. \eqref{eq:direct-density-app} and Eq.~\eqref{eq:update-density-app} give
\begin{equation}
    f_{t+1}(q)
    =
    \frac{
        f_0(q)\mathbf{1}_{\{q_0\leq q\leq b_{t-1}\}}
        \mathbf{1}_{\{q_0\leq q\leq b_t\}}
    }{
        \int_{q_0}^{b_t}
        f_0(x)\mathbf{1}_{\{q_0\leq x\leq b_{t-1}\}}\,dx
    } .
\end{equation}
Because \(b_t<b_{t-1}\), this reduces to
\begin{equation}
    f_{t+1}(q)
    =
    \frac{
        f_0(q)\mathbf{1}_{\{q_0\leq q\leq b_t\}}
    }{
        \int_{q_0}^{b_t}f_0(x)\,dx
    } .
    \label{eq:direct-density-next-app}
\end{equation}

Therefore, the next-period mean quality is
\begin{equation}
    \bar{q}_{t+1}
    =
    \int q f_{t+1}(q)\,dq
    =
    \frac{
        \int_{q_0}^{b_t} q f_0(q)\,dq
    }{
        \int_{q_0}^{b_t} f_0(q)\,dq
    } .
\end{equation}
This proves the equivalence between sequential updating and direct truncation along a monotone deterioration path.

\bibliographystyle{cas-model2-names}
\bibliography{cas-refs}

\end{document}